\documentclass[aps,prl,twocolumn,superscriptaddress,floatfix]{revtex4-2}
\usepackage{multirow}
\usepackage[latin9]{inputenc}
\usepackage{graphicx}
\usepackage{float,epsfig}
\usepackage{dcolumn}
\usepackage{bm}
\usepackage{graphicx}
\usepackage{amsmath,amssymb,amsthm}
\usepackage{subfigure}
\usepackage{color}
\usepackage[colorlinks=true,linkcolor=blue]{hyperref}

\usepackage{times}
\usepackage{multirow}

\usepackage{pifont}

\newcommand{\cR}{\mathcal{R}}

\newcommand{\p}{\mathcal{ P}}

\newcommand{\T}{\mathcal{ T}}

\newcommand{\bC}{{\bm C }}

\newcommand{\bR}{{\bm R }}

\newcommand{\bn}{{\bm n}}

\newcommand{\br}{{\bm r}}

\newcommand{\bk}{{\bm k}}

\newcommand{\bzeta}{{\bm \zeta}}
\newcommand{\bmeta}{{\bm \eta}}

\newcommand{\rev}[1]{\textcolor{black}{{#1}}}

\begin{document}  

\title{Two-dimensional nonlinear Thouless pumping of matter waves} 

\author{Qidong Fu}
\affiliation{School of Physics and Astronomy, Shanghai Jiao Tong University, Shanghai 200240, China}
 
\author{Peng Wang}
\affiliation{School of Physics and Astronomy, Shanghai Jiao Tong University, Shanghai 200240, China}
 
\author{Yaroslav V. Kartashov}
 \affiliation{Institute of Spectroscopy, Russian Academy of Sciences, Troitsk, Moscow Region, 108840, Russia}
 
\author{Vladimir V. Konotop}
 \affiliation{Departamento de F\'{i}sica and Centro de F\'{i}sica Te\'orica e Computacional, Faculdade de Ci\^encias, Universidade de Lisboa, Campo Grande, Ed. C8, Lisboa 1749-016, Portugal}
 
 \author{Fangwei Ye}
 \email[]{\textcolor{black}{Corresponding author: fangweiye@sjtu.edu.cn}}
\affiliation{School of Physics and Astronomy, Shanghai Jiao Tong University, Shanghai 200240, China}
 
\begin{abstract} 
 
We consider \rev{theoretically} the nonlinear quantized Thouless pumping of a Bose-Einstein condensate loaded in a two-dimensional dynamical optical lattices. We encountered three different scenarios of the pumping: quasi-linear one occurring for gradually dispersing wave packets, transport carried by a single two-dimensional soliton, and multi-soliton regime when initial wave packet splits into several solitons. The scenario to be realized depends on the number of atoms in the initial wave packet and on the strength of the two-body interactions. The magnitude and direction of the displacement of a wavepacket are determined by Chern numbers of the populated energy bands and by the inter-band transitions induced by two-body interactions. As a case example we explore a separable potential created by optical lattices whose constitutive sublattices undergo relative motion in the orthogonal directions. For such potentials, obeying parity-time symmetry, fractional Chern numbers, computed over half period of the evolution, acquire relevance. We focus mainly on solitonic scenarios, showing that one-soliton pumping occurs at relatively small as well as at sufficiently large amplitudes of the initial wavepacket, while at intermediate amplitudes the transport is multi-solitonic. We also describe peculiarities of the pumping characterized by two different commensurate periods of the modulations of the lattices in the orthogonal directions.

\end{abstract} 

\maketitle 

Since its discovery by Thouless~\cite{Thouless} the quantized transport was in the focus of extensive studies for electrons in condensed matter~\cite{NiuThoul,Niu,MiChaNiu2010}, in spinor systems~\cite{spin}, fermionic~\cite{NaToTa2016,TaCoFa2017,NaTaKe2021} and bosonic~\cite{LosZilAlBlo2020,LoSHa2018} atomic gasses, in photonic~\cite{KraLaRin2012, ZilHuaJo2018, CeWaShe2020} and acoustic~\cite{CheProPro2020} systems, and in plasmonic waveguides~\cite{FedQiLIK2020}. Originally the quantized transport (alias Thouless pumping) was studied in one-dimensional (1D) linear systems. Experimental observations of this phenomenon in optical and atomic systems have triggered investigations of the impact of the space dimension and of the nonlinearity on transport. 

One can distinguish two 2D implementations of the quantized transport. The first one, that has received the most attention so far, is motivated by the intimate relation between the quantized transport and different Hall effects. Such transport occurs along boundaries of finite 2D structures and is carried by the boundary modes~\cite{ZilHuaJo2018}. Being strongly confined to the boundaries, this mechanism is quasi-one-dimensional in the real space, although it may require the use of effective multidimensional spaces for its analytical description. The second implementation is the transport carried by bulk modes (in a formally unlimited potential). In addition to a quantized one-cycle displacement such transport is characterized by a quantized angle determining the direction of motion. The latter setting was implemented experimentally with a BEC in an optical superlattice~\cite{LosZilAlBlo2020} and in tilted moir\'e lattices~\cite{our-PNAS}. 2D Thouless pumping of ultracold fermions was studied in~\cite{MaTeKa}.

It was established theoretically~\cite{NaYoKa2018,NonliPump_our,JuRe,MoGruGo} and confirmed experimentally~\cite{Rechtsman}, that in 1D systems the nonlinearity (originated by two-body interactions in atomic gasses and by the Kerr effect in optical waveguides) can lead to the breakup of the pumping. On the other hand, the nonlinearity sustains robust transport against disorder~\cite{TanDasAn2016}, allows one to achieve inversion of the pumping direction ~\cite{NonliPump_our}, and enables fractional pumping~\cite{TaCoFa2017,HaDuKAm2019}). Quantized transport can also be induced by interactions~\cite{Kuno}.

1D Thouless pumping of large-amplitude wavepackets in Bose-Einstein condensates (BECs) with a negative scattering length occurs in a form of solitons, whose one-cycle displacement is determined by populations of several lowest bands varying due to inter-band tunneling and characterized by the {\em first} Chern numbers of the populated {\em linear} bands~\cite{NonliPump_our}. Thus, the nonlinear quantized transport inherits the topological properties of the underlying linear system, while the nonlinearity controls the displacement through the coupling of populated bands.

In this Letter, we aim at considering quantized transport in a 2D BEC with a negative scattering length in the physical realization that goes far beyond the \rev{finite} tight-binding model explored previously~\rev{\cite{JuRe}} and has features having no analogs in the \rev{continuous} 1D systems~\cite{NonliPump_our}. Properties of pumping \rev{that cannot be observed in the previous settings are as follows}. First, a wavepacket described by the 2D Gross-Pitaevskii equation (GPE) with attractive interactions may collapse if the number of atoms exceeds a specific value~\cite{collapse1,collapse2}. Second, while an optical lattice (OL) can in principle stabilize 2D solitons~\cite{BaiMalSal,MiMaLe}, it is not obvious {\em apriori} that such stabilization persists for high-amplitude wavepackets evolving in dynamical OLs.  Third, there is \rev{a threshold number of atoms required for soliton formation, i.e., for a solitonic transport.} Consequently, the continuous model considered here reveals novel scenarios of nonlinear quantized transport carried either by a quasi-linear wavepacket, by a \rev{dynamically stable} single soliton, or by several solitons emerging during the decay of the initial wavepacket \rev{due to instability induced by inter-band tunneling}. The nonlinearity determines which of these scenarios is realized, as well as the absolute value and direction of the one-cycle displacement. Furthermore, we report on peculiarities of pumping by lattices with different temporal periods in the orthogonal directions.

Consider a quasi-2D condensate described by the dimensionless GPE
\begin{align}
	\label{GPE}
	i\frac{\partial \Psi}{\partial t}=-\frac{1}{2}\nabla^2\Psi+[V^x(x,t)+V^y(y,t)]\Psi-|\Psi|^2\Psi.
\end{align}
Here $\nabla=(\partial_x,\partial_y)$, $\br=(x,y)$ is measured in the units of $d/\pi$ where $d$ is the smallest of the lattice constants, time $t$ is measured in the units of $md^2/(\pi^2\hbar)$, the amplitudes of the OL $V^\xi (\xi,t)= V^\xi (\xi+\pi,t)=V^\xi (\xi,t+T_\xi)$ (hereafter $\xi=x,y$), are given in the units of $md^2/(\pi\hbar)^2$, $T_{x,y}$ are temporal periods which can be different but commensurate. The $\Psi$ is normalized as ${N:=\int |\Psi|^2d\br=\mathcal{N}(\pi/2)^{1/2}|a_s|/a_\bot}$, where $\mathcal{N}$ is the total number of atoms, and $a_\bot$ is the condensate extension along the $z$ direction \rev{(e.g., for a $^7$Li BEC with $a_s\approx -1.43\,$nm in a trap with $a_\bot\approx 1\,\mu$m the norm  $N=1$ corresponds to $\mathcal{N}=558$ atoms)}. We assume that $T_x=nT_y$ where $n$ is an integer, i.e., $T=T_x$ is the
period of the 2D lattice.  

Due to separability of the potential in Eq.~(\ref{GPE}) one can introduce 1D Hamiltonians $H^\xi=-(1/2)\partial_\xi^2+V^\xi(\xi,t)$ and consider the eigenvalue problems $H^\xi\varphi_{\alpha_\xi k_\xi}^\xi=\varepsilon_{\alpha_\xi k_\xi}^\xi(t)\varphi_{\alpha_\xi k_\xi}^\xi$ for the Bloch functions $\varphi_{\alpha_\xi k_\xi}^\xi
=e^{ik_{\xi} \xi}u_{\alpha_\xi k_\xi}^\xi(\xi,t)$, where $u_{\alpha_{\xi} k_{\xi}}^\xi(\xi,t)=u_{\alpha_{\xi} k_{\xi}}^\xi(\xi+\pi,t)$, $k_\xi\in[-1,1)$, and $\alpha_\xi$ are the indices of the 1D bands. The spectrum of the full Hamiltonian $H=H^x+H^y$ is given by $\varepsilon_{\nu \bk}(t)=\varepsilon_{\alpha_x k_x}^{x}(t)+\varepsilon_{\alpha_y k_y}^{y}(t)$, where $\bk=(k_x,k_y)$ and $\nu$ is the index of 2D bands at $\bk={\bf 0}$ and $t=0$. We restrict the consideration to potentials for which $\varepsilon_{\nu_1 {\bf 0}}(0)<\varepsilon_{\nu_2 {\bf 0}}(0)$ if $\nu_1<\nu_2$. Such a choice implies one-to-one correspondence $\nu\leftrightarrow (\alpha_x,\alpha_y)$, with $\nu=1$ corresponding to $(\alpha_x,\alpha_y)=(1,1)$. If no crossing of the 1D bands occurs at $t>0$, the labeling of $\varepsilon_{\nu \bk}(t)$ by $\nu$ 
remains well defined at any $t$ and $\bk$. Now one can define Chern numbers 
of the 1D potentials:
\begin{align}
\label{Chern}
	C_{\alpha_\xi }^\xi(T_\xi)= \frac{i}{2\pi }\int_{0}^{T_\xi}\!\!dt\int_{-1}^{1}dk\left( \langle \partial_t u_{\alpha_\xi k}^\xi | \partial_{k} u_{\alpha_\xi k}^\xi\rangle
	\right. \nonumber \\ \left.
	-\langle  \partial_{k} u_{\alpha_\xi k}^\xi | \partial_t u_{\alpha_\xi k}^\xi\rangle \right).
\end{align}
(here $\langle f | g \rangle:=\int_{0}^{\pi} \bar{f}(\xi)g(\xi)d\xi$ and an overbar denotes complex conjugation). 

We characterize pumping by the displacement of the center of mass (c.m.) of the wave packet ${\bm {\cR}}(t)=(1/N)\int_{\mathbb{R}^2}\br|\Psi|^2d^2r$. Defining
2D Wannier functions: $W_{\nu }^{\bn}(\br,t)=w_{\alpha_x n_x}^x
(x,t) w_{\alpha_y n_y}^y (y,t)$ where $\bn=(n_x,n_y)$, $n_{x,y}$ are integers, and $w_{\alpha_\xi n_\xi}^\xi(\xi,t)
 ={({1}/{2})}\int_{-1}^1\varphi_{\alpha_\xi k}^\xi
(\xi,t)
e^{-i\pi n_\xi k}dk$ are 1D Wannier functions~\cite{Kohn,review_Wannier}, for   $N<N_{\rm cr}$, where $N_{\rm cr}$ is the critical number of atoms above which the initial wavepacket collapses, we expand
$
\Psi(\br, t) 
=\sqrt{N}\sum_{\nu=1}^\infty\sum_{\bn} a_{\nu }^{\bn} (t) W_{\nu }^{\bn}(\br,t). $
The  coefficients $a_{\nu }^{\bn} (t)$ are normalized $\sum_{\nu,\bn}|a_{\nu }^{\bn}|^2=1$. Using this expansion one can show that ${\bm {\cR}}(t) = \bR(t)+\bmeta(t)+\bzeta(t)$~\cite{supplemental}. Here  $\bR(t)=\left(X(t), Y(t)\right) =\sum_{\nu} \rho_{\nu}\left(X_{\alpha_x},Y_{\alpha_y}\right)$ describes dynamics of the c.m. due to adiabatic change of the potential, $\rho_\nu(t)=\sum_{\bn}|a_{\nu }^{\bn} (t)|^2$ is the relative population of the $\nu$-th band ($\sum_{\nu}\rho_\nu=1$), while  $X_{\alpha_x}(t)=\int |w_{\alpha_x 0}^x|^2 xdx$ and $Y_{\alpha_y}(t)=\int  |w_{\alpha_y 0}^y|^2ydy $ are independent displacements (also known as polarizations~\cite{Vander}) along the $x$ and $y$ directions. 
The   $\bmeta(t)=\pi \sum_{\bn}\eta^{\bn}\bn$, where 
$\eta^{\bn}=\sum_{\nu}|{a}_{\nu }^{\bn}|^2$, describes change of the c.m. position due to the tunneling between potential wells  
while $\bzeta(t)$  encompasses simultaneous spatial tunneling between minima of the potential accompanied by the inter-band transitions. Equations for $a_{\nu}^{\bn}(t)$, as well as
the explicit form of $\bzeta$ are given in~\cite{supplemental}.  
\rev{While dispersive spreading of a small-amplitude wave packet can be a non-adiabatic process occurring faster than the motion of a c.m., the pumping still can be observed in this case. For a separable potential such a limit becomes a superposition of independent pumping along $x-$ and $y-$directions determined by $X_{\alpha_x}$ and $Y_{\alpha_y}$. In weakly nonlinear 1D systems it was discussed in~\cite{NonliPump_our} (see also~\cite{supplemental})}. 

\rev{
Meantime, unlike in the cases of nonlinear pumping in a 1D GPE~\cite{NonliPump_our} and in a 2D finite discrete lattice~\cite{JuRe}, the domain of a quasi-linear pumping cannot be reduced to zero, because solitons of the 2D GPE do not bifurcate from the linear modes.} 
 For a soliton \rev{of (\ref{GPE})} to be created, the norm $N$  
must exceed some threshold value $N_{\rm th}$. Both $N_{\rm th}$ and $N_{\rm cr}$ 
depend on the specific parameters of the OL and in our case adiabatically vary in time. Therefore, these quantities are understood below as the maxima: $N_{\rm th, cr}=\max_t \{N_{\rm th, cr}(t)\}$, where $N_{\rm th, cr}(t)$ are the respective instantaneous values. At $N<N_{\rm th}$ the pumping is quasi-linear. At $N_{\rm th}<N<N_{\rm cr}$ the attractive nonlinearity can sustain {\em stable} 2D solitons~\cite{2Dsolitons1,2Dsolitons2}. It turns out that in solitonic regime two different scenarios occur: a single-soliton pumping and a multi-soliton pumping. In the former case, when only one soliton is created, the tunneling between neighbouring lattice minima is suppressed by the attractive inter-atomic interactions. If also $\bmeta(0)=\bzeta(0)={\bm 0}$, these quantities remain small at $t=T$ and tend to zero at $T\to\infty$~\cite{supplemental}. Below we consider wave packets  initially centered at ${\bm \cR}(0)={\bm 0}$, for which ${\bm {\cR}}(T)\approx \bR(T)$.  Then, since $X_{\alpha_x}(T_x)\approx \pi C_{\alpha_x}^x(T_x)$ and $Y_{\alpha_y}(T_y)\approx \pi C_{\alpha_y}^y(T_y)$, for quasi-linear and single-soliton scenarios one can estimate [${\bm\cR}_T:={\bm\cR}(T)$] 
\begin{align}
\label{main1}
    {\bm\cR}_T\approx\pi  \sum_{\nu} \rho_\nu(T)\bC_\nu, \quad \bC_\nu=\left(C_{\alpha_x}^{x}, nC_{\alpha_y}^{y}\right). 
\end{align}

Four comments are in order. First, at $n>1$ Eq.~(\ref{main1}) does not allow for equally accurate adiabatic description (formally valid for $T\to\infty$) of the displacements along $x$ and $y$ directions, since the latter undergoes $n$ cycles. Second, ${\bm\cR}_T$ includes the dependence on the band populations $\rho_\nu(T)$, that makes the nonlinear dynamics non-separable.
Third, as we show below, in some cases the formula (\ref{main1}) remains valid even at half-period $t=T/2$. Finally, transport may be accompanied by a non-adiabatic process of splitting of the initial wavepacket into several solitons. Although then $\bmeta(t)$ and $\bzeta(t)$ cannot be neglected, the approximation (\ref{main1}) remains meaningful for the interpretation of the physical results.

\begin{figure}
\includegraphics[width=\columnwidth]{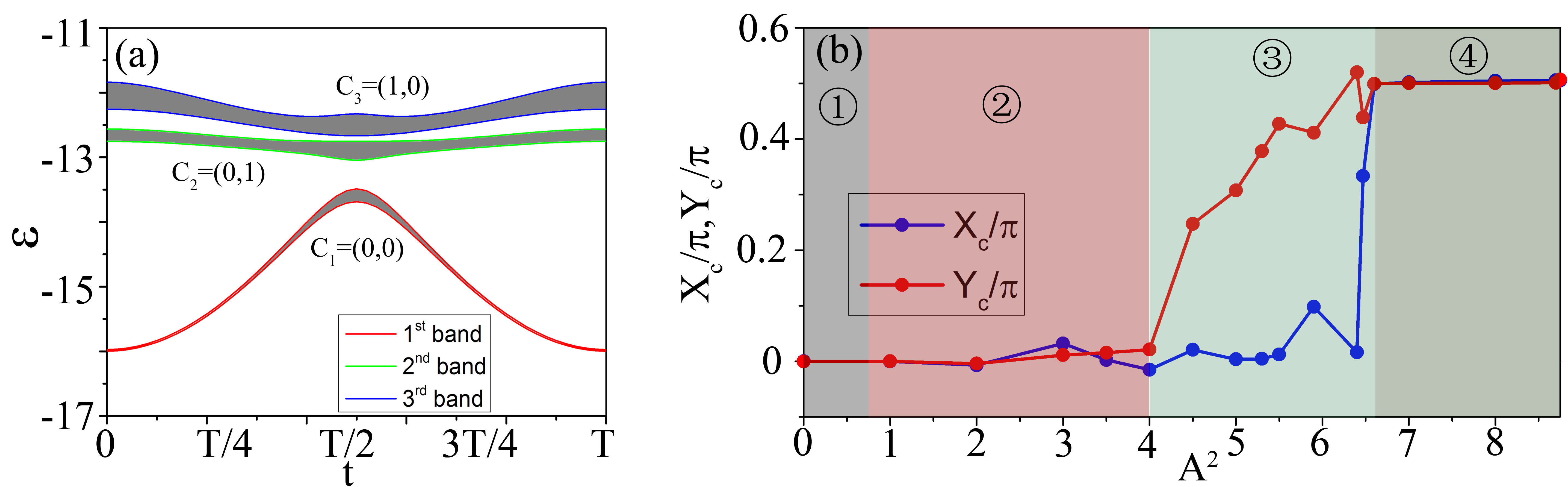}
\caption{(a) Evolution of 2D bands (\rev{grey} domains \rev{with boundaries shown by color lines}) over one cycle $T$ of variation \rev{of the potential (\ref{potential}) with}  $\mathfrak{V}_x=5$, $\mathfrak{V}_y=4$, $\mathfrak{v}_x=1.2$ and $\mathfrak{v}_y=2$. Here $\nu=1,\,2,\,3$ correspond to the pairs  $(1,1),\,(1,2),\,(2,1)$. The Chern numbers are indicated in the panel. (b) The  $(X_c,Y_c)$  obtained by the numerical simulations of Eq.~(\ref{GPE}) with $\Psi(\br,0)=A\exp(-r^2/0.7^2)$ 
{\em vs.} $A^2$, for $T=100\pi$. Domains indicated by the encircled numbers correspond to different pumping regimes. The dots indicate amplitudes for which numerical data are obtained.
	 }  
\label{fig1}
\end{figure}

To the illustration of the general picture described above we explore the potentials of the form \rev{($\xi=x,y$)}
\begin{align}
    \label{potential}
    V^\xi(\xi,t)=-\mathfrak{V}_\xi [\cos^2\xi+\mathfrak{v}_\xi\cos^2(2\xi-v_\xi t)],
\end{align}where $\mathfrak{V}_\xi$ and $\mathfrak{v}_\xi$ are positive amplitudes, and $v_\xi=\pi/T_\xi$ are the velocities of the relative motion of the constitutive lattices. Starting with the case of equal periods $T_{x,y}=T$ (and hence $v_{x,y}=v=\pi/T$), in Fig.~\ref{fig1}(a) we show an example of three lowest 2D bands of the above lattice and their Chern numbers. All results shown below are obtained for the Gaussian wavepacket $\Psi_{\rm in}=A\exp(-r^2/0.7^2)$ populating mainly the lowest band with
$\bC_1=(0,0)$ [Fig.~\ref{fig1}(a)]. Respectively, in the quasi-linear regime, corresponding to the parameter domain $\textcircled{1}$ in Figs.~\ref{fig1}(b), no transport is observed, as predicted by (\ref{main1}). \rev{Interestingly, this is a quantum effect since in the classical limit a corresponding particle in the potential $V^x+V^y$ would display a half-period shift in each direction (see~\cite{supplemental})}. 

\begin{figure*}[t]
\includegraphics[width=\textwidth]{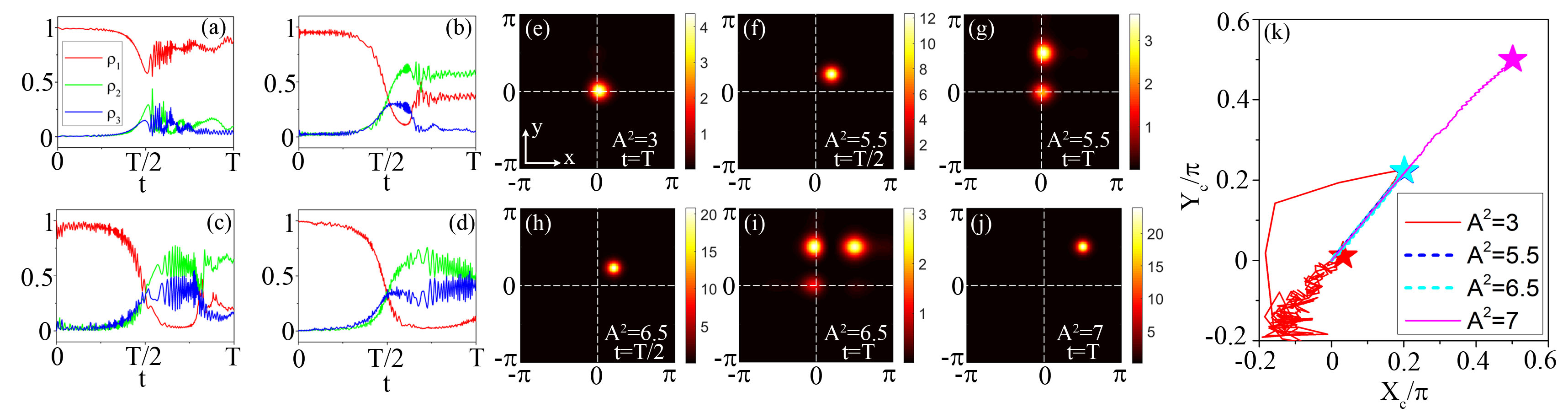}
\caption{Evolution of the population of the lowest bands for $A^2=3$ (a), $5.5$ (b),  $6.5$ (c), and $7$ (d). (e-j) Distributions $|\Psi|^2$ for the respective amplitudes at $t=T/2$ and $t=T$.
The crossings of dashed lines indicate initial positions of the wave-packet centers. In (k) the trajectory of c.m. for different initial amplitudes ended up in the locations shown by the stars at $t=T/2$ (dashed lines) and at $t=T$ (solid lines).   
}
\label{fig23}
\end{figure*}

The boundary between $\textcircled{1}$ and $\textcircled{2}$ domains at $A_{\rm th}\approx 0.87$ corresponds to $N=N_{\rm th}\approx 0.57$. Above this threshold a stable 2D soliton is created. For moderate nonlinearities $ 0.87\lesssim  A \lesssim 2$ in the entire domain $\textcircled{2}$, one still does not observe appreciable c.m. shifts at $t=T$. This can be understood by considering the populations of the higher bands during one-cycle evolution plotted in Fig.~\ref{fig23}(a). While weak nonlinearity couples the lowest (red line) band with the second (green line) and third (blue line) bands and tunneling does occur for $t<T$, the final distribution at $t=T$ is characterized by the dominant population of only the lowest band [$\rho_1(T)\approx 0.879$ in Fig.~\ref{fig23}(a)].  Respectively, as predicted by ${\bm \cR}_T\approx {\bm 0}$ in (\ref{main1}), at $t=T$ the soliton is found near the origin [Fig.~\ref{fig23}(e)].   Meantime    the trajectory at $0<t<T$ appears very irregular [red line in Fig.~\ref{fig23}(k)]. \rev{The observed long-time evolution is a numerical proof of the dynamical stability of a soliton.}  

Further increase of the input wavepacket amplitude, corresponding to the domain \textcircled{3} in Fig.~\ref{fig1}(b), enhances the inter-band tunneling, especially around $t\approx T/2$ [Fig.~\ref{fig23}(b) and Fig.~\ref{fig23}(c)]. At that instant the 1D lattices in (\ref{potential}) are $\pi-$shifted and the amplitudes of the potentials $V^\xi$ acquire their minimal values $\mathfrak{V}_\xi$. Inspecting band populations one observes that during first half period most of the atoms are in the lowest linear band, while right after the instant $T/2$ most of atoms tunnel to the second and third bands and new solitons are born. Now, the formula (\ref{main1}) is not applicable anymore at $t=T$, but it remains meaningful at $t=T/2$ (i.e., before the birth of new solitons) because the Chern numbers $C_{\alpha_\xi}^\xi (T_\xi)$, considered as functions of $T_\xi$, remain topological quantities upon the replacement $T_\xi\to T_\xi/2$. Indeed, the Hamiltonians $H^\xi$ with the potential (\ref{potential}) are parity-time symmetric, $[\p_\xi \T,H]=0$ where the parity operators $\p_\xi$ change $\xi\to-\xi$ and $\T$ is the  time reversal changing $t\to -t$ and performing complex conjugation. Bloch states of such Hamiltonians can also be chosen parity-time symmetric.
Then, using the $T_\xi$ periodicity of the potential $V^\xi(\xi,t)$, one can show~\cite{supplemental} that $C_{\alpha_\xi}^\xi (T_\xi/2)=C_{\alpha_\xi}^\xi (T_\xi)/2$, i.e. it is a (fractional) topological index and one can consider    transport for $0\leq t \leq T$. 
In Fig.~\ref{fig23}(f) and (h) we show density distributions at $t=T/2$ for the cases that at $t=T$ have different numbers of newborn solitons. In both cases the initial wavepackets have not undergo decay and have c.m. coordinates ${\bm \cR}_{T/2}=(0.20,0.22)\pi$. Moreover, the soliton trajectories in both cases are nearly the same: blue and cyan dashed lines in Fig.~\ref{fig23}(k) are indistinguishable. On the other hand, the half-period  displacements predicted by Eq.~(\ref{main1}) for band populations ($\rho_1,\rho_2,\rho_3)\approx (0.38,0.35,0.27)$ and $(0.38,0.30,0.32)$ corresponding to the panels (f) and (h), are ${\bm \cR}_{T/2}\approx (0.14,0.18) \pi$ and ${\bm \cR}_{T/2}\approx (0.16,0.15)\pi$.

Thus, while the approximation (\ref{main1}) qualitatively explains the observed quantized transport, it underestimates the magnitude of the half-period displacement. This can be explained by two factors. First, (\ref{main1}) uses the normalization to $N$ under the assumption that all atoms remain in soliton at $t=T/2$, but in reality a small fraction of atoms from the initial Gaussian wavepacket $\Psi_{\rm in}$ is always dispersed and does not contribute to soliton. 
Thus, in practice a soliton has smaller "mass" $N_{\rm s}$ than the total norm, $N_{\rm s}\lesssim N$, and a correction factor $f=N/N_{\rm s}$ should be considered: ${\bm \cR}\to f{\bm \cR}$. In our simulations $f\approx 1.1$. Second, the dynamics of band populations is characterized by initial nonadiabatic evolution during the process of soliton formation [small-scale oscillations at $t\ll T$ in Fig.~\ref{fig23}(b) and (c)]. Thus, effectively the adiabatic evolution "starts" at some $t_0>0$. To check this we considered $t_0\approx 0.02 T$ and computed $f{\bm \cR}_{0.52T}\approx(0.17,0.22)\pi$ and $f{\bm \cR}_{0.52T}=(0.21,0.23)\pi$ for the above cases, which is in excellent agreement with numerically obtained c.m. locations at $t=0.52T$: $(0.21,0.23)\pi$ and $(0.21,0.22)\pi$. 
 
Thus, the domain \textcircled{3} is characterized by two different regimes: single-soliton transport until approximately $T/2$, \rev{when the soliton loses its stability because of enhanced inter-band tunneling [see Fig.~\ref{fig23}(b)] and 2(c)}, followed by multi-soliton transport. The output density distributions at $t=T$ are illustrated in Fig.~\ref{fig23}(g) and (i). While the displacement at $T/2$ weakly depends on the input amplitude, at $t>T/2$ the dynamics crucially depends on $A$ determining the number of emerging solitons [cf. panels (g) and (f)]. At $t=T$ all newborn solitons are centered in the vicinity of the points $(m,n)\pi/2$ where $m,n=0,1$.

A counter-intuitive result is obtained upon further increase of the amplitude in the interval  $6.6 \lessapprox A^2\lessapprox 8.7$ [domain \textcircled{4} in Fig.~\ref{fig1}(b)].  The initial wavepacket evolves into a single large-amplitude soliton composed of atoms equally populating topological second and third bands. At $t=T$ this two-band 2D soliton is encountered at the approximate position $(\pi/2,\pi/2)$ [Fig.~\ref{fig23}(j)] in agreement with the formula (\ref{main1}).  
The right edge of the domain \textcircled{4} corresponds to the critical norm $N_{\rm cr}\approx 6.69$ [$A_{\rm cr}^2\approx 8.7$] above which BEC collapses. The collapse occurs at a time much smaller than $T$~\cite{supplemental}. When this happens the meanfield GPE does not describe all observed effects~\cite{Donley} requiring consideration of such effects as quantum fluctuations~\cite{DuSt,Yurovsky}.
 
\begin{figure}
	\includegraphics[width=\columnwidth]{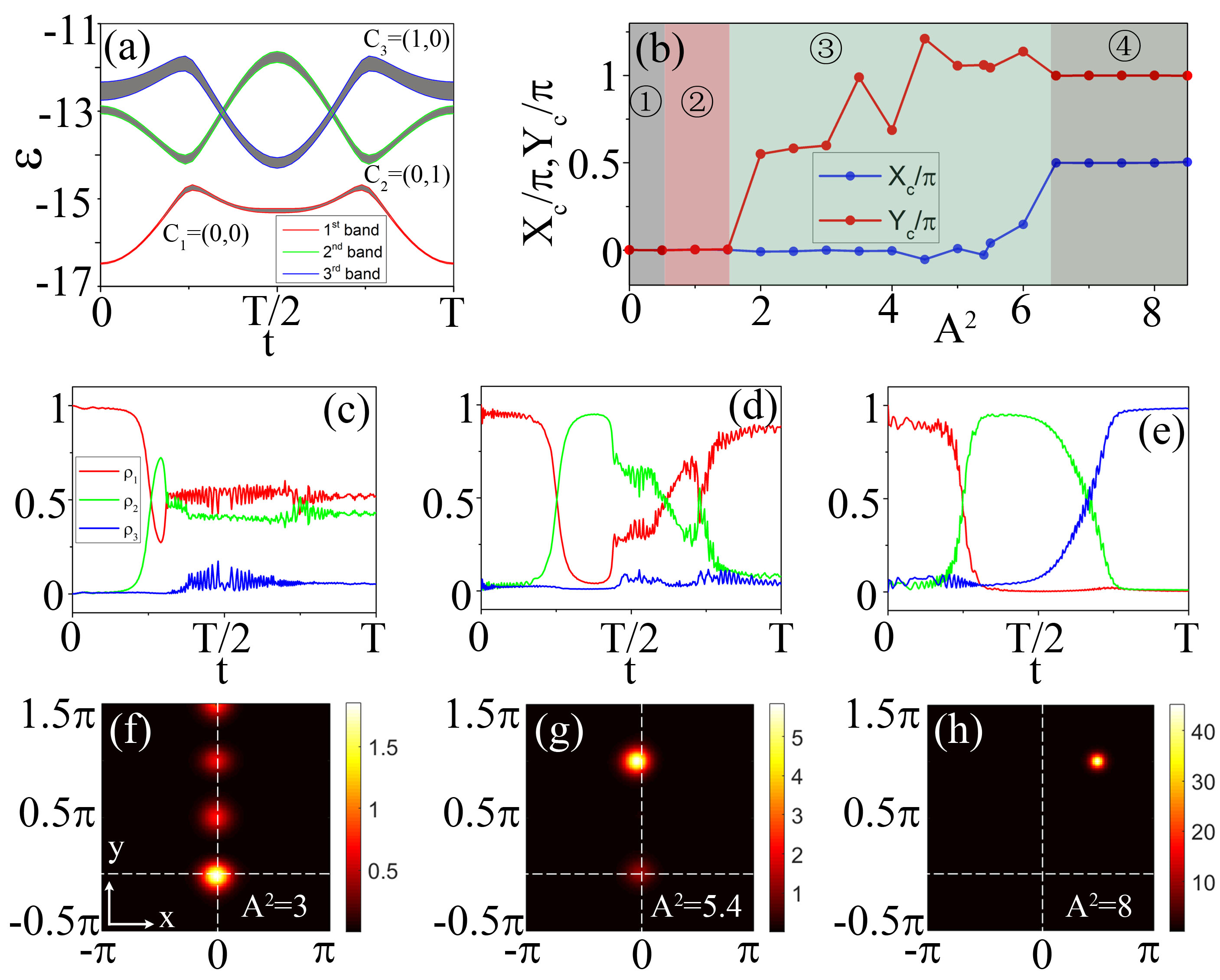}
	\caption{(a) Evolution of 2D bands of the potential (\ref{potential}) with $\mathfrak{V}_x=5$, $\mathfrak{v}_x=1.2$,  $\mathfrak{V}_y=4.2$, $\mathfrak{v}_y=2$ with $T=T_x=2T_y$. (b) C.m. coordinates at $t=T$.  Different colours correspond to scenarios shown in Fig.~\ref{fig1}(b).  Evolution of populations  $\rho_1$  (red), $\rho_2$ (green) and $\rho_3$ (blue) for $A^2=3$ (c), $5.4$ (d) and $8$ (e). (f-h) Distributions $|\Psi|^2$ at $t=T$ for the respective amplitudes. 
	The initial conditions are the same as in Fig.~\ref{fig1}. 
	}  
	\label{fig4}
\end{figure}

Nonlinear quantized transport in 2D allows one to consider non-equal periods of modulations along $x$ and $y$ directions.  To explore this possibility we consider the potential (\ref{potential}) with $T_x=2T_y=T$ ($v_y=2v_x$). Fig.~\ref{fig4} summarizes the results. For the chosen parameters the total second gap of the 2D lattice closes at certain instants of time [Fig.~\ref{fig4}(a)], e.g. around $t=T/2$ the second 2D band acquires larger energy than the third one. At certain times we also observe shrinking of the first finite gap resulting in the enhancement of the inter-band tunneling [Fig.~\ref{fig4}(c)-(e)]. Meantime, the gaps of the constitutive 1D lattices (not shown here) remain permanently open. Therefore, the definition of the Chern numbers (\ref{Chern}), indicated in Fig.~\ref{fig4}(a), remain valid. Likewise, the formula (\ref{main1}) with $n=2$ remains valid in the quasi-linear regime [domain  \textcircled{1} in Fig.~\ref{fig4}(b)] and in a single-soliton regime [domain~\textcircled{2} in Fig.~\ref{fig4}(b)] that are analogous to respective regimes for equal periods. Upon increase of the amplitude of the initial wavepacket we enter the domain, where the quantized transport is carried by several solitons. Now the first event of strong inter-band tunneling occurs approximately around $t\approx T/4$. For such instant of time one cannot identify topological indices. Therefore the formula  (\ref{main1}) can suggest only a qualitative interpretation of the results shown in Fig.\ref{fig4}(f),(g). Since all atoms remain in the lowest two bands during the one-cycle evolution the emerging solitons are located along the $y$-axis at nearly equidistant points $y_n\approx n\pi/2$ (with $n=0,1,...$).  

Finally, in the domain \textcircled{4} in Fig.~\ref{fig4}(b)%
~we again observe quantized transport carried by a large-amplitude soliton [Fig.~\ref{fig4}(h)]. The one-cycle displacement of the soliton, $(\pi/2,\pi)$ in Fig.~\ref{fig4}(h) can be understood using (\ref{main1}) if we consider it as one-cycle shift of the second band soliton $\pi \bC_2(T)$ over the one period of lattice in the $y$-direction, i.e., at $T/2$ where the only the second band is populated [green line in Fig.~\ref{fig4}(e)], and subsequent displacement of the third-band soliton $\pi \bC_3(T/2)$ [blue line in Fig.~\ref{fig4}(e)] over a half-period of the $x-$lattice.

We have described 2D nonlinear Thouless pumping of an attractive BEC in adiabatically varying \rev{commensurate} OL. The transport can occur in quasi-linear, single-soliton, or multi-soliton regimes. The scenario is determined by the dynamical Chern numbers of the linear bands and by the band populations affected by nonlinearity-induced inter-band tunneling.  Single-soliton transport occurs for relatively small and high-amplitude wavepackets. For  amplitudes between these limits, the transport is carried by several solitons born upon the decay of the initial wavepacket. The concept of 2D quantized transport remains meaningful also for non-equal commensurate temporal periods of the mutually orthogonal OLs. \rev{Nonlinear quantum transport in continuous lattices with incommensurate periods remains an open question.}

\acknowledgements

Q.F., P.W. and F.Y. acknowledge support from NSFC (No.91950120,11690033), Natural Science Foundation of Shanghai  (No.19ZR1424400), and Shanghai Outstanding Academic Leaders Plan (No. 20XD1402000). 
V.V.K. acknowledges financial support from the Portuguese Foundation for Science and Technology (FCT) under Contracts PTDC/FIS-OUT/3882/2020 and UIDB/00618/2020.


\begin{thebibliography}{0}%
\makeatletter
\providecommand \@ifxundefined [1]{%
 \@ifx{#1\undefined}
}%
\providecommand \@ifnum [1]{%
 \ifnum #1\expandafter \@firstoftwo
 \else \expandafter \@secondoftwo
 \fi
}%
\providecommand \@ifx [1]{%
 \ifx #1\expandafter \@firstoftwo
 \else \expandafter \@secondoftwo
 \fi
}%
\providecommand \natexlab [1]{#1}%
\providecommand \enquote  [1]{``#1''}%
\providecommand \bibnamefont  [1]{#1}%
\providecommand \bibfnamefont [1]{#1}%
\providecommand \citenamefont [1]{#1}%
\providecommand \href@noop [0]{\@secondoftwo}%
\providecommand \href [0]{\begingroup \@sanitize@url \@href}%
\providecommand \@href[1]{\@@startlink{#1}\@@href}%
\providecommand \@@href[1]{\endgroup#1\@@endlink}%
\providecommand \@sanitize@url [0]{\catcode `\\12\catcode `\$12\catcode
  `\&12\catcode `\#12\catcode `\^12\catcode `\_12\catcode `\%12\relax}%
\providecommand \@@startlink[1]{}%
\providecommand \@@endlink[0]{}%
\providecommand \url  [0]{\begingroup\@sanitize@url \@url }%
\providecommand \@url [1]{\endgroup\@href {#1}{\urlprefix }}%
\providecommand \urlprefix  [0]{URL }%
\providecommand \Eprint [0]{\href }%
\providecommand \doibase [0]{https://doi.org/}%
\providecommand \selectlanguage [0]{\@gobble}%
\providecommand \bibinfo  [0]{\@secondoftwo}%
\providecommand \bibfield  [0]{\@secondoftwo}%
\providecommand \translation [1]{[#1]}%
\providecommand \BibitemOpen [0]{}%
\providecommand \bibitemStop [0]{}%
\providecommand \bibitemNoStop [0]{.\EOS\space}%
\providecommand \EOS [0]{\spacefactor3000\relax}%
\providecommand \BibitemShut  [1]{\csname bibitem#1\endcsname}%
\let\auto@bib@innerbib\@empty
\end{thebibliography}%


\begin{thebibliography}{99}
	
	\bibitem{Thouless}   D. J. Thouless,  Quantization of particle transport, {Phys. Rev. B} {\bf 27}, 6083 (1983).
	
	
	\bibitem{NiuThoul} Q. Niu and D. J. Thouless, Quantised adiabatic charge transport in the presence of substrate disorder and many-body interaction, J. Phys. A: Math. Gen. {\bf 17} 2453 (1984).
	
	\bibitem{Niu} Q. Niu, Towards a Quantum Pump of Electric Charges, Phys. Rev. Lett. {\bf 64}, 1812 (1990). 
	
	\bibitem{MiChaNiu2010} D. Xiao, M.-C. Chang, and Q.  Niu,  Berry phase effects on electronic properties, Rev. Mod. Phys. {\bf 82}, 1959 (2010).
	
	\bibitem{spin} 
	W. Ma, L. Zhou, Q. Zhang, M. Li, C. Cheng, J. Geng, X. Rong, F. Shi, J. Gong, and J. Du, Experimental Observation of a Generalized Thouless Pump with a Single Spin, Phys. Rev. Lett. {\bf 120}, 120501 (2018).
	
	\bibitem{NaToTa2016} S. Nakajima, T. Tomita, S. Taie, T. Ichinose, H.  Ozawa, L. Wang, M. Troyer, and Y. Takahashi, Topological Thouless pumping of ultracold fermions.  Nat. Phys. {\bf 12}, 296 
	(2016).
	
	\bibitem{NaTaKe2021} S. Nakajima, N. Takei, K. Sakuma, Y. Kuno, P. Marra, and  Y. Takahashi,
	Competition and interplay between topology and quasi-periodic disorder in Thouless pumping of ultracold atoms. Nat. Phys.  {\bf 17}, 844 (2021).
	
	\bibitem{TaCoFa2017}
	L. Taddia, E. Cornfeld, D. Rossini, L. Mazza, E. Sela, and R. Fazio, Topological Fractional Pumping with Alkaline-Earth-Like Atoms in Synthetic Lattices, Phys. Rev. Lett. {\bf 118}, 230402 (2017).
	
	\bibitem{LosZilAlBlo2020} M. Lohse, S. Schweizer, O. Zilberberg, M. Aidelsburger, and I. Bloch,  A Thouless quantum pump with ultracold bosonic atoms in an optical superlattice,  Nat. Phys. {\bf 12}, 350 -- 354 (2016).
	
	\bibitem{LoSHa2018} M. Lohse, C. Schweizer, H. M. Price, O. Zilberberg, and I. Bloch, Exploring 4D quantum Hall physics with a 2D topological charge pump, Nature {\bf 553}, 55
	(2018). 
	
	
 	\bibitem{KraLaRin2012} Y. E. Kraus,  Y. Lahini, Z. Ringe, M. Verbin, and  O. Zilberberg, 
	Topological States and Adiabatic Pumping in Quasicrystals,  { Phys Rev. Lett.} {\bf 109}, 106402 (2012).
	
	\bibitem{ZilHuaJo2018} O. Zilberberg, S. Huang, J. Guglielmon, M. Wang, K. P. Chen, Y. E. Kraus, and M. C. Rechtsman, Photonic topological boundary pumping as a probe of 4D quantum Hall physics,  Nature, {\bf 553}, 59 (2018).
	
	\bibitem{CeWaShe2020} A. Cerjan, M. Wang,  S. Huang, K. P. Chen, and M. Rechtsman, Thouless pumping in disordered photonic systems, Light: Science \& Applications {\bf 9}, 178 (2020).
	
	
	\bibitem{CheProPro2020} 
	W. Cheng, E. Prodan, and C. Prodan, 
	Experimental Demonstration of Dynamic Topological Pumping across Incommensurate Bilayered Acoustic Metamaterials, Phys. Rev. Lett. {\bf 125}, 224301 (2020).
	
	
	\bibitem{FedQiLIK2020} Z. Fedorova, H. Qiu, S. Linden, and  J. Kroha,  
	Observation of topological transport quantization by dissipation in fast Thouless pumps,
	{Nat. Comm.} {\bf 11}, 3758 (2020).
	
	
	%
	
	 
	
	
	 \bibitem{our-PNAS} P. Wang,  Q. Fu, R. Peng, Y. V. Kartashov,  L. Torner,  V. V. Konotop,  and  F. Ye, Two-dimensional Thouless pumping of light in photonic moir\'e lattices (submitted).
	
	
	%
	\bibitem{MaTeKa}  F. Matsuda, M. Tezuka,  and N. Kawakami,  Two-Dimensional Thouless Pumping of Ultracold Fermions in Obliquely Introduced Optical Superlattice, J. Phys. Soc. Jpn. {\bf 890}, 114708 (2020).
	%
	
\bibitem{NaYoKa2018} M. Nakagawa, T. Yoshida,  R. Peters,  and N. Kawakami, Breakdown of topological Thouless pumping in the strongly interacting regime,
	Phys. Rev. B {\bf 98}, 115147 (2018).
	
\bibitem{NonliPump_our} Q. Fu, P. Wang, Y. V. Kartashov, V. V. Konotop, and F. Ye, Nonlinear Thouless pumping: solitons and transport breakdown, Phys. Rev. Lett. {\bf 128}, 154101 (2022).
	
\bibitem{JuRe} M. J\"urgensen and M. C. Rechtsman, Chern Number Governs Soliton Motion in Nonlinear Thouless Pumps, Phys. Rev. Lett. {\bf 128}, 113901 (2022).

\bibitem{MoGruGo} N. Mostaan, F. Grusdt, and N. Goldman, Quantized transport of solitons in nonlinear Thouless pumpings: From Wannier drags to topological polarons  https://arxiv.org/abs/2110.08696	
	
\bibitem{Rechtsman} M. J\"urgensen, S. Mukherjee, and M. C. Rechtsman, Quantized nonlinear Thouless Pumping, Nature {\bf 596}, 63
		(2021).   	
		
\bibitem{TanDasAn2016} J. Tangpanitanon, V. M. Bastidas, S. Al-Assam, P. Roushan, D. Jaksch, and D. G. Angelakis, Topological Pumping of Photons in Nonlinear Resonator Arrays, Phys. Rev. Lett. {\bf 117}, 213603 (2016).		
	
\bibitem{HaDuKAm2019} T Haug, R. Dumke, L.-C. Kwek, and L. Amico, Topological pumping in Aharonov -- Bohm rings, Comm. Phys. {\bf 2}, 127 (2019).

	
\bibitem{Kuno} Y. Kuno and Y. Hatsugai, Interaction-induced topological charge pump, Phys. Rev. Res. {\bf 2}, 042024(R) (2020).	

\bibitem{collapse1} 
	L. Berg\'e, Wave collapse in physics: Principles and applications to light and plasma waves, Phys. Rep. {\bf 303}, 259 (1998).
	
\bibitem{collapse2} G. Fibich, The Nonlinear Schr\"odinger Equation: Singular Solutions and Optical Collapse (Springer, Heidelberg, 2015).

\bibitem{BaiMalSal} B. B. Baizakov, B. A. Malomed, and M. Salerno, Multidimensional solitons in a low-dimensional periodic potential, Phys. Rev. A {\bf 70}, 053613 (2004).

\bibitem{MiMaLe} D. Mihalache, D. Mazilu, F. Lederer, Y. V. Kartashov, L.-C. Crasovan, and L. Torner, Stable three-dimensional spatiotemporal solitons in a two-dimensional photonic lattice, Phys. Rev. E {\bf 70}, 055603(R) (2004).

\bibitem{Kohn} W. Kohn, Analytic Properties of Bloch Waves and Wannier Functions, Phys. Rev. {\bf 115}, 809 (1959).
	
\bibitem{review_Wannier} Marzari, N.,   Mostof, A. A.,   Yates,  J. R.,  Souza, I. and Vanderbilt, D., Maximally localized Wannier functions: Theory and applications,  {Rev. Mod. Phys.} {\bf 84}, 1419 (2012).

\bibitem{supplemental} See derivation in the Supplemental material, which includes Refs. [19,20,30,33].	
\bibitem{AKKS} G. L. Alfimov, P. G. Kevrekidis, V. V. Konotop, and M. Salerno, Wannier functions analysis of the nonlinear Schr\"odinger equation with a periodic potential, Phys. Rev. E {\bf 66}, 046608 (2002).


\bibitem{Vander} R. D. King-Smith and D. Vanderbilt, Theory of polarization of crystalline solids. Phys. Rev. B {\bf 47}, 1651
(1993).


 \bibitem{BluKon}	 Yu. V. Bludov and V. V. Konotop, Surface modes and breathers in finite arrays of nonlinear waveguides, Phys. Rev. E {\bf 76}, 046604 (2007).
 
	
\bibitem{2Dsolitons1} B. B. Baizakov, B. A. Malomed, and M. Salerno, Multidimensional solitons in periodic potentials, Europhys. Lett. {\bf 63}, 642 (2003).

\bibitem{2Dsolitons2} B. B. Baizakov, B. A. Malomed, and M. Salerno, Multidimensional solitons in a low-dimensional periodic potential, Phys. Rev. A {\bf 70}, 053613 (2004). 

\bibitem{Donley} E. A. Donley, N. R. Claussen, S. L. Cornish, J. L. Roberts, E. A. Cornell and C. E. Wieman. Dynamics of collapsing and exploding Bose-Einstein condensates, Nature {\bf 412}, 299 (2001). 	

\bibitem{DuSt} R. A. Duine and H. T. C. Stoof, Explosion of a Collapsing Bose-Einstein Condensate, Phys. Rev. Lett. {\bf 86}, 2204 (2001).

\bibitem{Yurovsky} V. A. Yurovsky, Quantum effects on dynamics of instabilities in Bose-Einstein condensates, Phys. Rev. A {\bf 65} 033605 (2002)

 
 

 
	
\end{thebibliography}
\end{document}